\documentclass[final]{aa}  

\usepackage{graphicx}
\usepackage{txfonts}
\usepackage{xargs}              % Use more than one optional parameter in a new commands
\usepackage{orcidlink}          % allow ORCID ids
\usepackage{hyperref}
\hypersetup{
  colorlinks   = true, %Colours links instead of ugly boxes
  urlcolor     = blue, %Colour for external hyperlinks
  linkcolor    = blue, %Colour of internal links
  citecolor    = blue  %Colour of citations
}
\usepackage{natbib}
\usepackage{xcolor}
\usepackage{float}
        
\begin{document}

\title{Observational insights into Sr~{\sc i} 4607\,\AA\ scattering polarization\\ with DKIST/ViSP}

\author{ 
         Franziska Zeuner \orcidlink{0000-0002-3594-2247} \inst{1}\thanks{Corresponding author}              
\and     Ernest Alsina Ballester  \orcidlink{0000-0001-9095-9685} \inst{2}$^,$\inst{3} 
\and     Luca Belluzzi \orcidlink{0000-0002-8775-0132} \inst{1}$^,$\inst{4} 
\and     Roberto Casini \orcidlink{0000-0001-6990-513X} \inst{5} 
\and     David M. Harrington \orcidlink{0000-0002-3215-7155} \inst{6}
\and     Tanaus\'{u} del Pino Alem\'{a}n \orcidlink{0000-0003-1465-5692} \inst{2}$^,$\inst{3}  
\and     Javier Trujillo Bueno \orcidlink{0000-0001-5131-4139} \inst{2,3,7}\thanks{Affiliate scientist of the National Center for Atmospheric Research, Boulder, U.S.A.}
}

   \institute{
   $^1$ Istituto ricerche solari Aldo e Cele Daccò (IRSOL), Faculty of Informatics, Università della Svizzera italiana, 6605 Locarno, Switzerland\\
   $^2$ Instituto de Astrof\'{i}sica de Canarias, 38205 La Laguna, Tenerife, Spain\\
   $^3$ Departamento de Astrof\'{i}sica, Facultad de F\'{i}sica, Universidad de La Laguna, 38206 La Laguna, Tenerife, Spain\\ 
   $^4$ Euler Institute, Faculty of Informatics, Università della Svizzera italiana, 6900 Lugano, Switzerland\\
   $^5$ High Altitude Observatory, National Center for Atmospheric Research, P. O. Box 3000, Boulder, CO 80307-3000, USA\\
   $^6$ National Solar Observatory, Makawao, Hawaii, United States\\
   $^7$ Consejo Superior de Investigaciones Cientif\'{i}cas, Spain
   }

   \date{Received August 25, 2025, Accepted February 27, 2026}

% \abstract{}{}{}{}{}
% 5 {} token are mandatory
 
  \abstract
   {
    Scattering polarization signals in the Sr~{\sc i} 4607\,\AA\ spectral line are among the strongest originating from the solar photosphere, offering a powerful diagnostic of tangled magnetic fields in the 3–300\,G range via the Hanle effect. However, measuring them with sub-arcsec resolution remains a significant challenge, because their detection demands exceptionally precise and accurate observational techniques.
    }
   {
   We analyze spatially resolved quiet Sun observations of these signals  performed with the Visible Spectropolarimeter (ViSP) at the Daniel K. Inouye Solar Telescope (DKIST), and to identify its current observational limits.
   } 
   {
   We present high-resolution, high-precision spectropolarimetric observations in a spectral window including the Sr~{\sc i} 4607\,\AA\ line at various limb distances. 
   We applied consistent instrumental corrections across all spectral lines, enabling the adjacent lines to serve as reliable references.
   }
   {
   Close to the limb, the signal-to-noise is high and we confirm that the center-to-limb variation of scattering polarization is compatible with previous studies.
   At a limb distance of $\mu = 0.74$, the signal-to-noise is low but sufficient in the total linear polarization map to directly reveal sub-arcsec structures in the Sr~{\sc i} line for the first time, which can be attributed to scattering polarization. 
   Disk-center measurements are still dominated by the noise related to the current limitations of the observational setup.
   }
   {By combining high spatio-temporal and spectral resolution with exceptional polarimetric precision, DKIST enables measurements of solar photospheric scattering polarization at fine scales. 
   These advances open new possibilities for using scattering polarization as a diagnostic tool for detecting tangled magnetic fields at small spatial scales, offering deeper insights into the solar small-scale dynamo.
   However, current signal-to-noise limitations still hinder direct detection of disk-center scattering polarization and must be addressed before further progress can be made.
   }

   \keywords{Polarization; Scattering; Methods: observational; Sun: photosphere; Techniques: polarimetric
               }

   \maketitle

%%%%%%%%%%%%%%%%%%%%%%%%%%%%%%%%%%%%%%%%%%%%%%%%%%%%%%%%%%%%%%
\section{Introduction}
\label{sec:intro}
%%%%%%%%%%%%%%%%%%%%%%%%%%%%%%%%%%%%%%%%%%%%%%%%%%%%%%%%%%%%%%

Magnetic fields in the solar atmosphere drive a wide range of dynamic processes, from coronal heating to solar wind acceleration. 
Even the quietest regions of the solar photosphere contribute to this magnetic complexity \citep[e.g.,][]{Bellot2019}, acting as reservoirs of magnetic energy that can feed into or amplify large atmospheric phenomena. 
Despite their importance, the exact mechanisms by which photospheric magnetic fields influence the upper layers of the solar atmosphere remain poorly understood. 
Their small-scale structure, potentially tangled at very fine spatial scales, is commonly referred to as the \emph{turbulent} magnetic field component of the photosphere. 
While simulations suggest that these tangled fields are concentrated in intergranular lanes due to small-scale dynamo action \citep{Vogler2007,Rempel2014,Rempel2023}, direct observational constraints remain limited \cite[see, however, discussion by][]{TrujilloBueno2006}.
Standard Zeeman-effect diagnostics, though powerful for detecting large-scale fields, struggle to capture any signatures of these unresolved magnetic structures in polarization. 
Nevertheless, alternative Zeeman-based approaches have been explored: \citet{TrellesArjona2021} demonstrated that multi-line inversions of intensity spectra can be used to infer an average magnetic field in the deep photosphere, providing indirect constraints on the hidden quiet-Sun magnetism.
As a complementary and more direct probe, scattering polarization—shaped by the magnetic field via the Hanle effect—has emerged as a crucial diagnostics of the unresolved magnetism of the photosphere \citep{Stenflo1982,Faurobert-Scholl1993,Faurobert-Scholl1995,TrujilloBueno2004, Snik2010, Kleint2010}. 
A key step toward understanding these tangled magnetic fields is the ability to detect the predicted spatially resolved maps of photospheric scattering polarization, resulting from computationally expensive numerical simulations \citep{TrujilloBueno2007,DelPinoAleman2018}.
In the past few years, measurements of the Sr~{\sc i} 4607\,\AA\ line have validated such theoretical predictions of disk-center scattering polarization, using statistical de-noising techniques to enhance weak signals \citep{Zeuner2020,Zeuner2024}. 
However, going from these statistical insights to spatially resolved observations of scattering polarization still remains a formidable challenge, and was only recently achieved by \citet{Zeuner2025}. 

Recently, \citet{DelPinoAleman2021} provided approximate predictions for the expected signals with the Visible Spectro-Polarimeter \citep[ViSP;][]{DeWijn2022} at the Daniel K. Inouye Solar Telescope \citep[DKIST;][]{Rimmele2020,Rimmele2022}, which has the world’s largest aperture for solar observations, located in Hawaii, USA.
A crucial finding was that with ViSP/DKIST a sufficient polarization contrast at sub-arcsec scales is preserved in spite of the relatively long integration times (up to a minute\footnote{Even with a five-minute integration, the polarization contrast remains considerable.}) required for high polarimetric sensitivity. 
Building on these advances, we present high-resolution spectropolarimetric observations taken with this instrument, aiming to resolve the spatial variation of scattering polarization in quiet Sun regions at several limb distances.
By combining high spatio-temporal resolution with sufficient polarimetric sensitivity (better than 5$\cdot 10^{-3}$ of the intensity), our observations directly reveal small-scale structures in solar scattering polarization, confirming some theoretical predictions and providing new observational insights.
These results lay the groundwork for future studies of unresolved magnetic fields in quiet Sun regions. 
At the same time, we highlight current instrumental limitations of ViSP that must be addressed to fully exploit its potential for scattering polarization diagnostics. 

%%%%%%%%%%%%%%%%%%%%%%%%%%%%%%%%%%%%%%%%%%%%%%%%%%%%%%%%%%%%%%
\section{Observation and data reduction}
\label{sec:data}
%%%%%%%%%%%%%%%%%%%%%%%%%%%%%%%%%%%%%%%%%%%%%%%%%%%%%%%%%%%%%%

In this section we describe the main characteristics of the observed data (Section~\ref{sec:observations}) and the reduction procedure (Section~\ref{sec:reduction}) to obtain the polarization maps.

%-----------------------------------------
\subsection{Observations}
\label{sec:observations}
%-----------------------------------------

On September 1 and 3 2024, observations of the quiet Sun were acquired with DKIST's ViSP during its second observing cycle. 
ViSP’s second arm recorded a spectral region of about 8\,\AA\ including the Sr~{\sc i} 4607\,\AA\ line. 
This is the first DKIST campaign in which this spectral line was included. 
The observing sequence was designed to perform high-resolution deep scans of the solar disk.

Each scan consisted of 20 slit positions, with a slit width of 0\farcs0536, matching the 0\farcs0534 step size between positions. 
The total map size is 61\farcs8$\times$1\farcs1. Each map was repeated either two or three times, with a cadence of 10~min. 
Typical integration times for ViSP are of the order of a few seconds per slit position.
Here, it was 30~seconds, ensuring sufficient exposure to detect scattering polarization in Sr~{\sc i}.  
However, the camera duty cycle was only 46\%, meaning that the detector was actively exposed for just 13.8~seconds\footnote{Ten modulation states accumulated 115 times with 12\,ms exposure per state.} out of the 30-second integration time.
The  spatial sampling is 0\farcs024~\,pixel$^{-1}$ along the slit, while the spectral sampling is about 9.1\,m\AA\,pixel$^{-1}$.
The slit was always aligned with the parallactic angle in order to minimize differential atmospheric refraction. 
Consequently, although the slit had a fixed orientation on the sky, its angle relative to the local solar limb changed depending on where on the solar disk we observed. 

We selected three data sets (each containing up to three scans of the same region), one at the limb, one at disk center, and one at an intermediate limb distance, based on the best intensity image quality.
Thanks to particularly favorable seeing conditions during the limb observations,
all three scans taken there were used, whereas for the intermediate and disk-center positions, only the best single scan was selected for further analysis.
The limb distance of each data set is defined as $\mu = \cos(\theta)$ (according to the ViSP header), where $\theta$ is the heliocentric angle, and refers to the center of the slit. 
At the solar limb\footnote{Product ID/Dataset ID: \href{https://dkist.data.nso.edu/product/L1-LWINP}{L1-LWINP/AJZRR}}, we have $\mu = 0.15$; at the intermediate position\footnote{Product ID/Dataset ID: \href{https://dkist.data.nso.edu/product/L1-WTXPV}{L1-WTXPV/BMNRV}}, $\mu = 0.74$; and at disk center\footnote{Product ID/Dataset ID: \href{https://dkist.data.nso.edu/product/L1-MIYID}{L1-MIYID/AVXNW}}, $\mu = 0.998$.
The original reference direction for $+Q$ is aligned perpendicular to the central meridian of the Sun.

The observed fields of view (FoV) covered quiet Sun regions, except at the limb, where a pore was used to lock the adaptive optics (AO) system. 
The AO remained locked for 100\% of the analyzed scan steps. 
The image quality, as inferred from the intensity root-mean-square contrast of $\sim$10\%  in the core of the Sr~{\sc i} line (see \citealt{Zeuner2025} for details), indicates good and stable seeing conditions during the observations.

%-----------------------------------------
\subsection{Data reduction}
\label{sec:reduction}
%-----------------------------------------

Our data reduction procedure followed the standard data reduction pipeline of ViSP, complemented by additional processing steps to address residual issues. 
Below, we briefly describe these steps.

The ViSP calibration and reduction pipeline was applied to the data, including dark current removal, spatial and spectral flat-fielding, geometric calibration for dual-beam alignment, and polarimetric calibration.\footnote{\url{https://docs.dkist.nso.edu/projects/visp/en/v2.16.7/l0_to_l1_visp.html}} 
Dual-beam reduces seeing-induced cross-talk and increases the signal-to-noise ratio.
We recall that polarimetrically calibrating ViSP involves a system calibration of DKIST \citep[see][]{Harrington2023}. 
Polarimetric efficiencies for the linear polarization states are between 0.5 and 0.6. 
For further details we refer the reader to the data quality reports published with the data sets to be found on the DKIST Data Center Archive.

The spatially averaged continuum polarization amplitude at disk center is expected to be negligible. 
However, spurious Stokes $Q$, $U$, or $V$ signals at continuum wavelengths can arise from seeing-induced or instrumental cross-talk with Stokes $I$. 
To reduce this instrumental $I \rightarrow Q, U, V$ cross-talk, we applied only the continuum-based correction step of the empirical method described by \citet{Sanchez1992}. 
Although this correction is not physically meaningful \citep[see][]{Jaeggli2022}, we did not perform a full Mueller matrix calibration because the polarization levels in our data are very low and the DKIST polarimetric system calibration is sufficiently accurate for the present analysis.

\begin{figure}[ht]
    \includegraphics[width=0.5\textwidth]{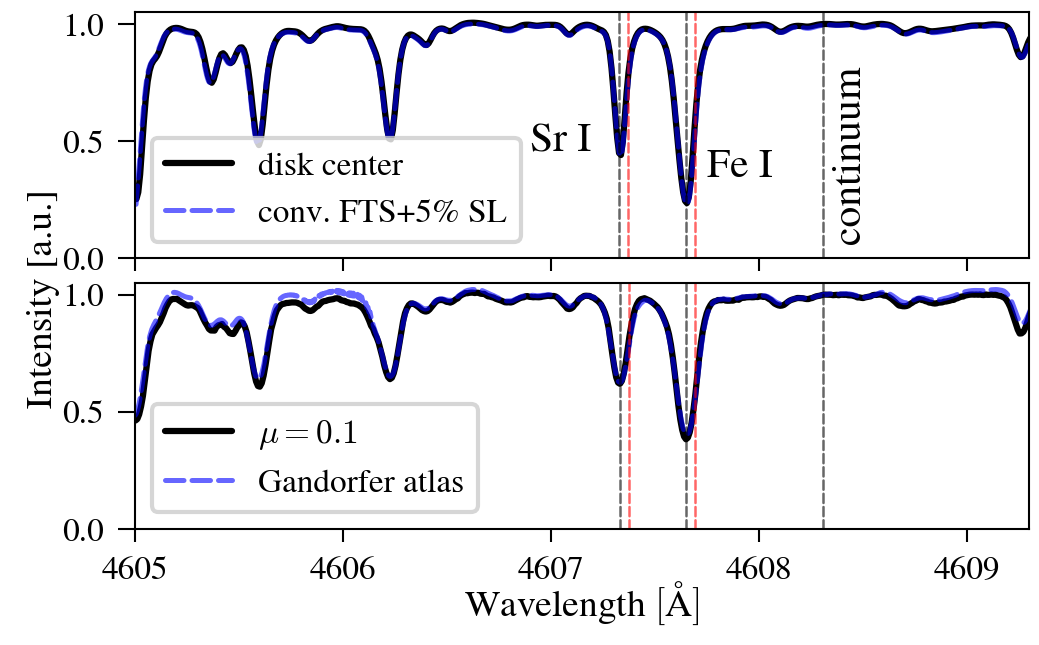}
    \caption{
       Intensity spectra at disk center and close to the limb compared to the FTS atlas \citep[][convolved with a 10 m\AA\ Gaussian and adding 5\% stray light]{Neckel1999} and the Second Solar Spectrum atlas \citep{Gandorfer2002}, respectively. Vertical dashed black lines indicate the spectral positions of the line cores (Sr~{\sc i} and  Fe~{\sc i}) and the selected continuum position. Vertical red lines indicate the wavelength positions of the red wing where Stokes $V/I$ is plotted later in the paper. We normalized each spectrum to the continuum. 
    }
    \label{fig:intensity_spectra}
\end{figure}

We identified a continuum spectral point (see Fig.~\ref{fig:intensity_spectra}) and used a spectral window of about 150\,m\AA\ to perform the cross-talk correction. 
Although this region is not a true continuum because of the presence of several weak spectral lines, it was carefully selected to avoid the cores of any significant lines, thereby minimizing contamination and ensuring a stable reference for correction.

The observed spectral window lacks intrinsically unpolarized telluric lines, which limits direct validation of this correction. 
Additionally, some of the observed solar regions were far from the disk center, weakening the assumption of an unpolarized continuum necessary for the ad-hoc method. 
Instead of direct validation, we convince ourselves \textit{a posteriori} of the efficiency of our full data processing pipeline by comparing the fractional linear polarization in the spectral lines of our data with the Second Solar Spectrum atlas \citep{Gandorfer2002}, as discussed later in the text.

The reference direction for the linear polarization was rotated to make $+Q$ parallel to the nearest solar limb. 
To perform this rotation, we calculated the angle between the solar rotation axis and the line connecting each spatial pixel to the center of the solar disk.
This angle was derived from the helioprojective-cartesian coordinates available in the data headers and using the small-angle approximation \citep{Thompson2006}.
It is noted in the data caveats that the accuracy of the reported coordinates is relatively low.\footnote{\url{https://nso.atlassian.net/servicedesk/customer/article/3481272321}}
Because of the limited observed FoV, it could not be reliably correlated with context images \citep[e.g., from the Visible Context Imager, VBI;][]{Woger2021} 
for improving the coordinate accuracy of the $\mu = 0.74$ data set.
The estimated accuracy of these coordinates is about 7\arcsec, based on our own limb observations (see next section). % confirmed with helpdesk
However, this number is likely a lower limit and the true error may be closer to 10\arcsec.
Limited coordinate accuracy can introduce residual rotation between $Q$ and $U$.
This rotation is negligible near the solar limb, where the relative error of the angle is low, but becomes increasingly relevant closer to disk center, where small angular errors have a larger impact.

For consistency with scattering polarization observations presented in the literature, the polarization profiles were normalized to the corresponding intensity profile at each pixel.
Throughout this paper, we therefore present and discuss fractional polarization values, though for simplicity we refer to them simply as "polarization."

The absolute polarization level measured with the current ViSP observations has not yet been fully validated, hindering the determination of the continuum polarization level \citep[e.g.,][]{Fluri1999,TrujilloBueno2009}. 
To avoid potential misinterpretations, we forced the continuum in all polarization states to zero. 
This was achieved by subtracting the mean polarization value in a spectral window of about 150\,m\AA\ around the identified continuum point (to reduce statistical noise) from all wavelength points, for each polarized Stokes parameter and each spatial pixel separately.\footnote{Note that this step might introduce minor cross-talk from Stokes $I$ near the limb, where some continuum polarization is expected; however, we do not find any evidence of this in our data.} 
Despite these corrections, small residual artifacts remained in the spectral lines.
These are mainly visible in regions where the scattering polarization signal is intrinsically weak, such as near disk center, where even relatively minor artifacts become noticeable.
Based on their spectral shape, we attribute these artifacts to a combination of known instrumental errors.
One contributing factor is the intensity imbalance between the two orthogonally polarized beams in the dual-beam set-up, which was about a factor of 1.7. 
This imbalance arises primarily from the grating polarization and, to a lesser extent, the imperfect contrast of the analyzing polarizer. 
Such imbalances amplify errors in one beam relative to the other in dual-beam data reduction, as previously noted by \citet{Casini2012a} and \citet{DeWijn2022}, and are known to reduce dual-beam polarimetric efficiency.

Although all calibration steps are incorporated in the data reduction pipeline and the measured polarimetric efficiencies fall within expected ranges, some level of uncertainty remains inherent to any polarimetric data processing.
ViSP was known to exhibit internal ghosting\footnote{See data set caveats: \url{https://nso.atlassian.net/wiki/spaces/DDCHD/pages/3481272321/ViSP+Data+Set+Caveats}} from multiple reflections, which vary with arm and wavelength. 
In our analysis, we control for potential ghost-related artifacts by performing a comparative assessment using the Fe~{\sc i} line.

For the subsequent analysis of intensity and linear polarization Stokes maps, we focus on three fixed wavelength points: the core of Sr~{\sc i}, the core of Fe~{\sc i}\footnote{In reality, this line is a blend of two Fe~{\sc i} transitions.}, and the continuum. 
These points, indicated with black dashed vertical lines in Fig.~\ref{fig:intensity_spectra}, are determined from the spatial mean intensity of each data set, and are used consistently throughout our analysis rather than selecting, for example, the maximum polarization value. 
The latter is ill-defined in the presence of noise and can vary unpredictably across pixels.

An alternative approach is to track the Doppler-shifted spectral line core for each pixel. 
While this method did not alter our qualitative conclusions, we observed a slightly different noise distribution, shifted and broadened toward higher values. 
We attribute this to the fact that, in the Doppler-shifted approach, polarization values are sampled from various detector pixels on the ViSP sCMOS camera, each with its own noise characteristics. 
In contrast, the fixed-wavelength method uses a consistent pixel across the dataset, minimizing pixel-to-pixel noise variation and avoids assumptions on the yet unconstrained spectral shape of the Sr~{\sc i} line polarization at high spatial resolution.
While this approach may not fully exploit the spectral coherence of the data, it provides a conservative and reproducible reference for comparing future observations and simulations.
For simplicity we will refer to the specific wavelengths as Sr~{\sc i}, Fe~{\sc i}, and the continuum.
To ensure the reliability of our findings in the following sections, all Sr~{\sc i} signals are consistently compared to reference signals in the neighboring Fe~{\sc i} line and continuum. 
If not stated otherwise, the intensity is normalized arbitrarily.

Since the Stokes $V/I$ signals due to the longitudinal Zeeman effect in the absence of velocities are expected to be small in the line core, we decided to show the maps of $V/I$ in the red wing of the spectral lines.
The offset from the line cores is 45.5\,m\AA, and the position is indicated by the vertical dashed red 
lines in Fig.~\ref{fig:intensity_spectra}.
 
While the line depths of the intensity spectra at disk center differ by approximately 20\% from the FTS atlas by \citet{Neckel1999}, the agreement improves substantially when the FTS profiles are convolved with a Gaussian of standard deviation 10 m\AA\ and supplemented with a 5\% stray-light contribution\footnote{Based on the intensity profile, we estimate less than 1.5\% scattered light at 15\arcsec\ outside the limb as one contribution to the stray-light.}.
Near the limb, the ViSP spectra already show good agreement with the limb atlas by \citet{Gandorfer2002}, see Fig.~\ref{fig:intensity_spectra}.
Since the FTS atlas is effectively free of spectral stray light, this moderately good agreement suggests that spectral stray light in ViSP is low but not negligible at this wavelength. 
Given that DKIST and ViSP were designed for high-sensitivity observations of faint coronal and prominence structures \citep{DeWijn2022}, the residual stray light may originate from known internal ghosting in this optical arm.

In contrast, the comparison with the limb atlas by \citet{Gandorfer2002} is less straightforward. 
The line depths in the atlas are shaped by several factors, such as limb distance, spectral resolution, and stray light, which makes it difficult to determine the individual contribution of each.
Consequently, the close agreement between the line depths of ViSP and the limb atlas may be coincidental.

%

%%%%%%%%%%%%%%%%%%%%%%%%%%%%%%%%%%%%%%%%%%%%%%%%%%%%%%%%%%%%%%
\section{Results}
\label{sec:results}
%%%%%%%%%%%%%%%%%%%%%%%%%%%%%%%%%%%%%%%%%%%%%%%%%%%%%%%%%%%%%%

In this section, we present the observed and processed data as described in the previous section. 
We present full Stokes spectrograms at disk center in Section~\ref{sec:results_center}.
From the limb dataset, we present full Stokes spatial maps and the center-to-limb variation (CLV) of the intensity and linear polarization in Section~\ref{sec:results_limb}.
Finally, we present full Stokes spatial maps at limb distance $\mu=0.74$ in Section~\ref{sec:results_disk}. For the latter, we analyze the spatial distribution of the total linear polarization with respect to the photospheric structures given by the continuum intensity and line-of-sight (LOS) velocity. Finally, we estimate the spatial signal-to-noise distribution.

%-----------------------------------------
\subsection{Disk center}
\label{sec:results_center}
\begin{figure}[ht]
    \includegraphics[width=0.5\textwidth]{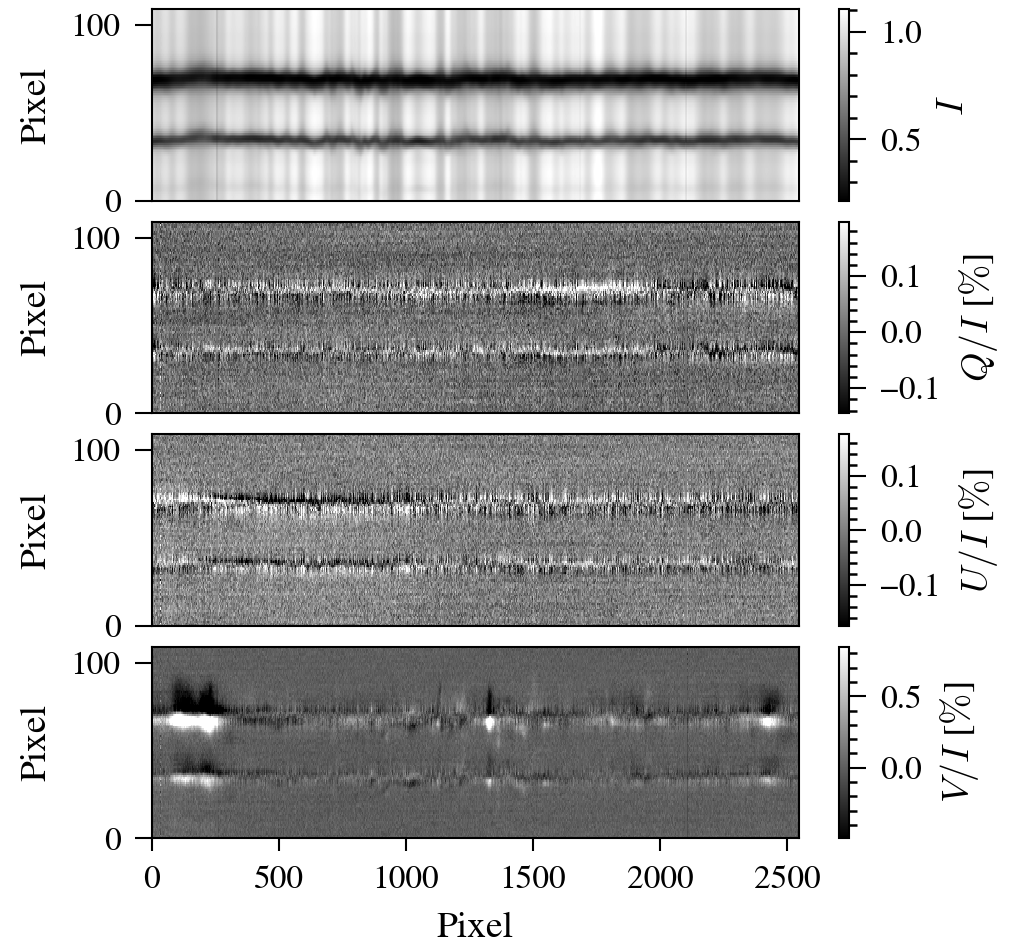}
    \caption{
       Full Stokes spectrograms of the first scan and fifth slit position at disk center, clipped to the spectral region of Sr~{\sc i} (bottom spectral line) and  Fe~{\sc i} (top spectral line). 
    }
    \label{fig:center_spectrum}
\end{figure}

%-----------------------------------------
Figure~\ref{fig:center_spectrum} shows the spectrograms in Stokes $I$, $Q/I$, $U/I$, and $V/I$ at disk center ($\mu = 0.998$) including both the Sr~{\sc i} and the neighboring Fe~{\sc i} lines. 
As expected, the $V/I$ spectra show the characteristic weak longitudinal Zeeman signals.
In contrast, the linear polarization signals exhibit weak but systematic structures. 
In both Sr~{\sc i} and Fe~{\sc i}, the $Q/I$ and $U/I$ spectra display $V$-like signatures, with one positive and one negative lobe (i.e., antisymmetric features) across the line profiles. 
These cannot be attributed to cross-talk from $V/I$ into $Q/I$ or $U/I$, as there is no significant correlation between these patterns and the circular polarization signal.
Given the magnetically relatively quiet conditions (as inferred from the low $V/I$ amplitudes), we attribute these signatures to instrumental effects. 
They are not consistent with transverse Zeeman effect signatures or intensity cross-talk.
As pointed out in Section~\ref{sec:reduction}, the likely origin are instrumental limitations, which may leave residual polarization artifacts in the data. 
Since these non-solar features dominate the linear polarization, we did not analyze this data set further.

%

%-----------------------------------------
\subsection{Limb}
\label{sec:results_limb}
%-----------------------------------------
The data set at the limb contains three scans with similar image quality. We investigate low spatial resolution spectra and the center-to-limb variation at high spatial resolution in Sections \ref{sec:results_limb_spectra} and \ref{sec:results_limb_scans}, respectively.

%-----------------------------------------
\subsubsection{Limb spectra}
\label{sec:results_limb_spectra}
\begin{figure}[ht]
    \includegraphics[width=0.5\textwidth]{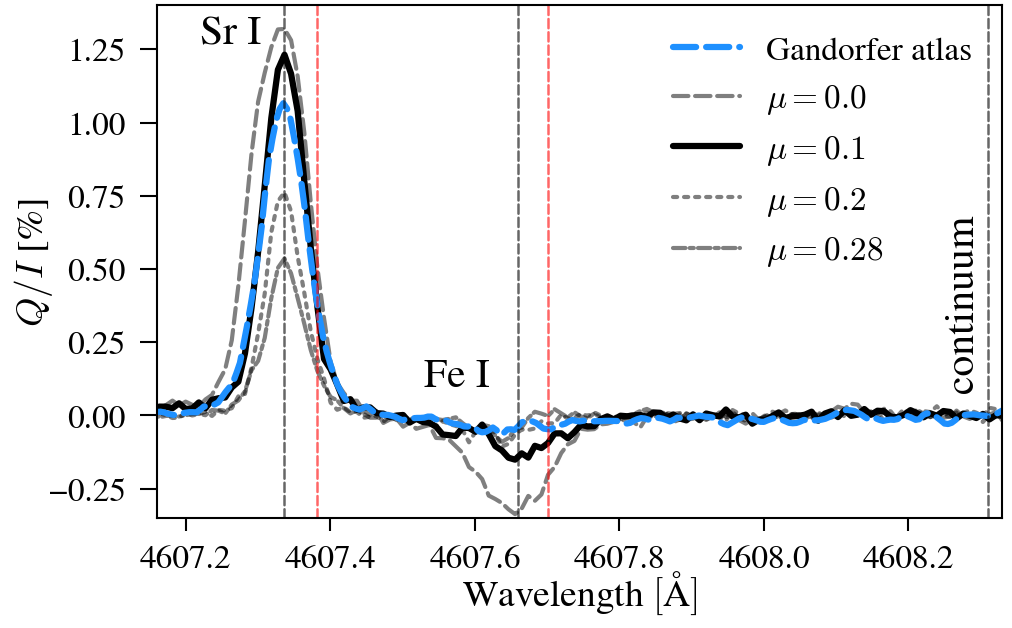}
    \caption{
       Linearly polarized spectra (obtained as explained in the text) at different limb positions $\mu$ (see legend). For reference, the data from the Second Solar Spectrum atlas \citep{Gandorfer2002} is also shown.
    }
    \label{fig:clv_q}
\end{figure}

%-----------------------------------------
To analyze the $Q/I$ spectra, we average all three scans and scan positions, which results in a low spatio-temporal resolution in the direction perpendicular to the slit, specifically in the radial direction.
We show the individual limb scans in Sr~{\sc i} in Appendix~\ref{sec:appendixA}.
From the continuum intensity distribution along the slit, which includes the limb because the slit is perpendicular to it, we determine the limb distance for each pixel along the slit.\footnote{We define  the limb as the pixel with the largest continuum intensity gradient along the slit, that is, in radial direction.}
The linear polarization spectra at several limb distances are shown in Fig.~\ref{fig:clv_q}.
In terms of spectral shape, the linear polarization signal in Sr~{\sc i} exhibits the characteristic single-lobed profile expected from scattering polarization and according to theoretical calculations. 
For lines that present scattering polarization, including the Sr~{\sc i} line, the linear polarization signal is highly sensitive to the limb distance, with polarization amplitudes increasing towards the limb. 
The maximum linear polarization at the extreme limb ($\mu=0.0$) is 1.27\%, which is significantly lower than what was reported by \citet[][up to 2\%]{Stenflo1997a}.
The figure also shows the linear polarization reported in the Second Solar Spectrum atlas by \citet{Gandorfer2002}, where we removed the continuum polarization for compatibility with our data. 
Their measured linear polarization parallel to the limb is slightly lower than what we find in our data set.
Their stated limb distance is $\mu=0.1$, but with limited accuracy, as the observations of \citet{Gandorfer2002} were obtained before active limb tracking became available at IRSOL.
Note that the spectral resolution of the atlas is 5\,m\AA\ and therefore higher than in the ViSP data. 
Therefore, small discrepancies in the linear polarization amplitude are expected.
Similar to the atlas, the neighboring Fe~{\sc i} line slightly depolarizes the continuum \citep{Stenflo1997a}, although at the extreme limb the depolarization is more pronounced.

%-----------------------------------------
\subsubsection{Spatial Stokes maps at the limb}
\label{sec:results_limb_scans}
%-----------------------------------------
%
\begin{figure*}[ht]
    \includegraphics[width=17cm]{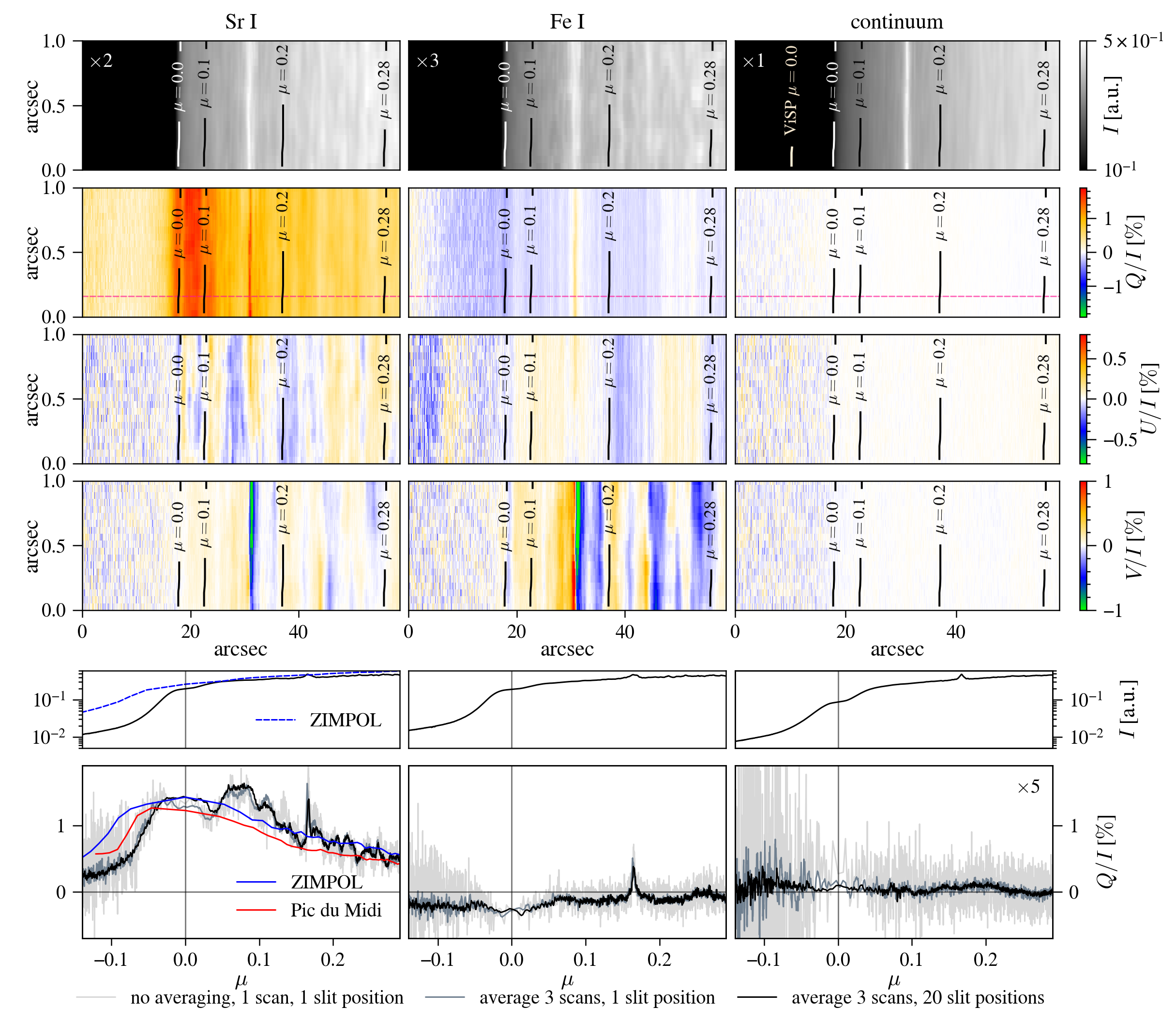}
    \caption{
     Four upper rows: Full Stokes maps for Sr~{\sc i} (left column), Fe~{\sc i} (central column), and the continuum (right column), averaged over all three scans and spatially binned to 0\farcs1$\times$0\farcs1 sampling. The intensity and linear polarization is plotted for the spectral line core wavelength positions, while the circular polarization is taken at a small wavelength offset in the red wing (see Fig.~\ref{fig:center_spectrum} for the offset). 
     Each Stokes parameter has the same scale shown on the right. The intensity is scaled by the factor given in the upper left of the intensity panel.
     Limb distance contours are given. The limb ($\mu=0$) is given by the inflection point of the continuum intensity.
     We also show the approximate position of $\mu=0$ determined from the ViSP coordinates in the continuum intensity image.
     Bottom rows: Center-to-limb variation (CLV) for the intensity and linear polarization parallel to the limb. Note that the continuum polarization is scaled by a factor of 5.
     Reference low spatio-temporal resolution Sr~{\sc i} data \citep{Zeuner2022,Malherbe2025} are shown in dark blue and red; see text for details.
     Note that the intensity is plotted on a logarithmic scale.
    }
    \label{fig:limb_images}
\end{figure*}
Figure~\ref{fig:limb_images} presents spatial Stokes maps of Sr~{\sc i}, Fe~{\sc i}, and the continuum (top panels), along with the CLV of intensity and $Q/I$ (bottom panels). 
The maps are averaged over all three available scans, which look similar in intensity and $Q/I$, representing a temporal integration of 30 minutes. 
This reduces statistical noise, and makes our results more easily comparable with previous studies. 
However, linear polarization fine structures that are clearly visible in the single scans of Sr~{\sc i} (see Appendix~\ref{sec:appendixA}) are lost when averaging. 
Due to the limited scan width relative to the slit length, the aspect ratio between the horizontal and vertical axes is strongly distorted, approximately 1:60. 
Iso-contours of $\mu$ are overplotted, where $\mu = 0$ is given by the inflection point of the continuum intensity profile along the limb direction.
The nominal solar limb position from ViSP’s coordinates given in the headers is also indicated in the continuum intensity panel, showing an offset of approximately 7\arcsec.

The on-disk Stokes maps in the continuum are free from any features, which gives us confidence that intensity cross-talk was sufficiently removed during the data reduction process.
The bright point at $\mu=0.18$ induces noticeable transverse Zeeman signals in $Q/I$ and longitudinal Zeeman signals in $V/I$ for the spectral lines. 
The $Q/I$ maps in Sr~{\sc i} and Fe~{\sc i} show the CLV behavior already discussed in the previous section. 
Interestingly, some linear polarization is visible beyond the limb position. 
We will discuss this aspect further below.

The $U/I$ maps in Sr~{\sc i} and Fe~{\sc i} show both positive and negative signals. 
However, they are only slightly larger in Sr~{\sc i} than in Fe~{\sc i}.
This could be due to temporal cancellation, because the single scans exhibit significantly stronger signals in Sr~{\sc i}.
However, it is difficult to determine the extent to which the Sr~{\sc i} $U/I$ signal arises from radiative transfer-induced polarization-produced by scattering of anisotropic radiation and by velocity gradients even in the absence of magnetic fields \citep[see][]{DelPinoAleman2018} -or from Hanle rotation \citep{Zeuner2022}, 
noting that  a contribution from the transverse Zeeman effect cannot be excluded.
The only way to distinguish between the Hanle and Zeeman effects would be to look at the spectra.
However, only some spectra in the FoV have an unambiguous single-lobe profiles in Sr~{\sc i}.
Although likely small, residual rotation between $Q/I$ and $U/I$ cannot be excluded completely, as discussed in  Section~\ref{sec:reduction}.
Given these challenges, we leave a further investigation and interpretation of the $Q/I$ and $U/I$ signals in terms of the Hanle effect for a future work.

Comparing the CLV of the Sr~{\sc i} intensity shown in the second-last row of Fig.~\ref{fig:limb_images} to the one shown in the ZIMPOL data taken at the Gregory-Coudé telescope of the IRSOL observatory \citep[0.45\,m aperture;][]{Zeuner2022}\footnote{Data is publicly available on ZENODO: \url{https://zenodo.org/records/8087406}.}, one can notice that the limb in DKIST is significantly sharper. 
This likely results from DKIST’s very low stray light levels (it was designed as a coronograph) and its adaptive optics system, which stabilizes the limb position far more effectively than is possible at IRSOL.
In the bottom row of Fig.~\ref{fig:limb_images}, we report the CLV of $Q/I$ in the core of Sr~{\sc i} from our analyzed ViSP dataset, ZIMPOL, and a low spatial resolution observation taken at Pic du Midi \citep[continuum polarization is subtracted;][]{Malherbe2025}.\footnote{The refractor was designed to be polarization free. Data is publicly available on HAL: \url{https://cnrs.hal.science/hal-05076008}.}
For ViSP, we show the CLV at three levels of spatial averaging: a single scan at a single slit position (no averaging), all scans averaged at that slit position, and an average over all scans and slit positions. 
The selected slit position is marked by the horizontal dashed pink line in the $Q/I$ image panels. 
All three ViSP observations show similar CLV trends. 
Compared to ViSP and ZIMPOL, the Pic du Midi amplitudes are slightly lower. 
A notable feature in all data sets is a local minimum in linear polarization very close to the limb in Sr~{\sc i}. 
The $Q/I$ signal drops sharply just outside the limb, a behavior also observed in molecular lines presented by \citet{Milic2012a}. 
As shown by the authors, this steep decline can be attributed to the sphericity of the solar atmosphere.
Remarkably, the general trend of the spatially-averaged $Q/I$ CLV (black curve) is highly compatible with ZIMPOL data \citep{Zeuner2022}, especially at the extreme limb ($\mu=0$).
However, the small-scale fluctuations are much more pronounced for ViSP, which is probably a result of the low spatio-temporal resolution of the ZIMPOL measurement. 
Also, the ZIMPOL $Q/I$ CLV shows scattering polarization signals further beyond of the limb, probably due to the point-spread function already affecting the intensity.

%-----------------------------------------
\subsection{Spatial Stokes maps at $\mu=0.74$}
\label{sec:results_disk}
%-----------------------------------------

From the intensity image of this observation, a spatial resolution of 0\farcs2 was determined by \citet{Zeuner2025}.
To improve the signal-to-noise (S/N) while preserving spatial information, we thus applied a spatial binning of 0\farcs1. 
This results in a final spatial sampling of 0\farcs1$\times$0\farcs1. 
In Appendix~\ref{sec:vbi}, a VBI context image in the 4500 \AA\ continuum\footnote{Product ID/Dataset ID: \href{https://dkist.data.nso.edu/product/L1-EBYNQ}{L1-EBYNQ/BUQAXN}} is provided for illustrative purposes, noting that the limited FoV of ViSP does not allow accurate co-alignment.

\subsubsection{Noise estimation}

Before the binning, the spatial standard deviation (which we call noise) in the continuum linear polarization is $\sigma$=0.032\% at a single spectral point.
After binning, the noise level in the continuum for the linear polarization is approximately $\sigma_b\approx$\,0.014\%.
This value is slightly higher than the expected 0.011\% derived from the unbinned data, assuming only statistical photon noise is present. 
The discrepancy suggests additional systematic errors may be present in the spatial dimension of the camera.
To account for these systematic effects when estimating the line-core noise from the continuum's noise, we apply a correction factor larger than one, ensuring that the true noise level is not underestimated. 
This factor reflects the observed discrepancy, which can reach up to nearly a factor of two for large binning values (used only to test the noise-scaling behavior), between the measured continuum noise and the photon-noise expectation.
Its empirical justification is discussed in Sect.\ref{sec:spatial_polarization}.

For pure photon noise, the standard deviation in Sr~{\sc i} 
should scale with $\sqrt{l}$, where $l$ is the continuum-to-line-depth-ratio. 
Averaged over the full field of view, we find $\langle \sqrt{l} \rangle \approx 1.4$ for Sr~{\sc i}. 
Including a worst-case correction factor of two, we obtain for the binned linear polarization an expected noise level of
$\sigma_{{\mathrm Sr}, b}=2\langle\sqrt{l}\rangle \,\sigma_b=0.04\%$.

To estimate the noise in the total linear polarization defined as \mbox{$P_{\mathrm L} = \left[\left(Q/I\right)^2 + \left(U/I\right)^2\right]^{1/2}$} in Sr~{\sc i} and Fe~{\sc i} from the continuum, we note that if the underlying distribution is normal, $P_\mathrm{L}$ follows a half-normal distribution. 
However, we find that the observed distribution $P_\mathrm{L, cont}$ in the continuum already deviates from this form: the ratio of the standard deviation $\sigma^{{P}}$ to the mean $\langle P_\mathrm{L, cont}\rangle$ is 0.5 instead of the theoretical approximately 0.8. 
This again points to the presence of systematic errors, implying that $\sigma^{{P}}$ is underestimated relative to the mean of the distribution.

We therefore construct a zero-signal reference distribution, i.e., a synthetic distribution representing the noise-only case, $P_\mathrm{L, Sr, zero} =  2\sqrt{l}\, P_\mathrm{L, cont}$, and define the noise value as $\sigma^{P}_{\mathrm Sr}=\langle P_\mathrm{L, Sr, zero} \rangle$. 
By using the mean of this distribution rather than its standard deviation, we obtain a conservative upper limit for the noise in the line core, since systematic tails in the distribution make the standard deviation an unreliable estimator in our case.
An analogous definition is applied to Fe~{\sc i}.
For the binned case, we use the binned continuum distribution $P_{\mathrm{L, cont}, b}$ as the starting point for constructing the zero-signal reference.
For the binned total linear polarization, this yields $\sigma^{P}_{{\mathrm Sr}, b}\approx0.11\%$.

\begin{figure*}[ht]
\centering
\includegraphics[width=17cm]{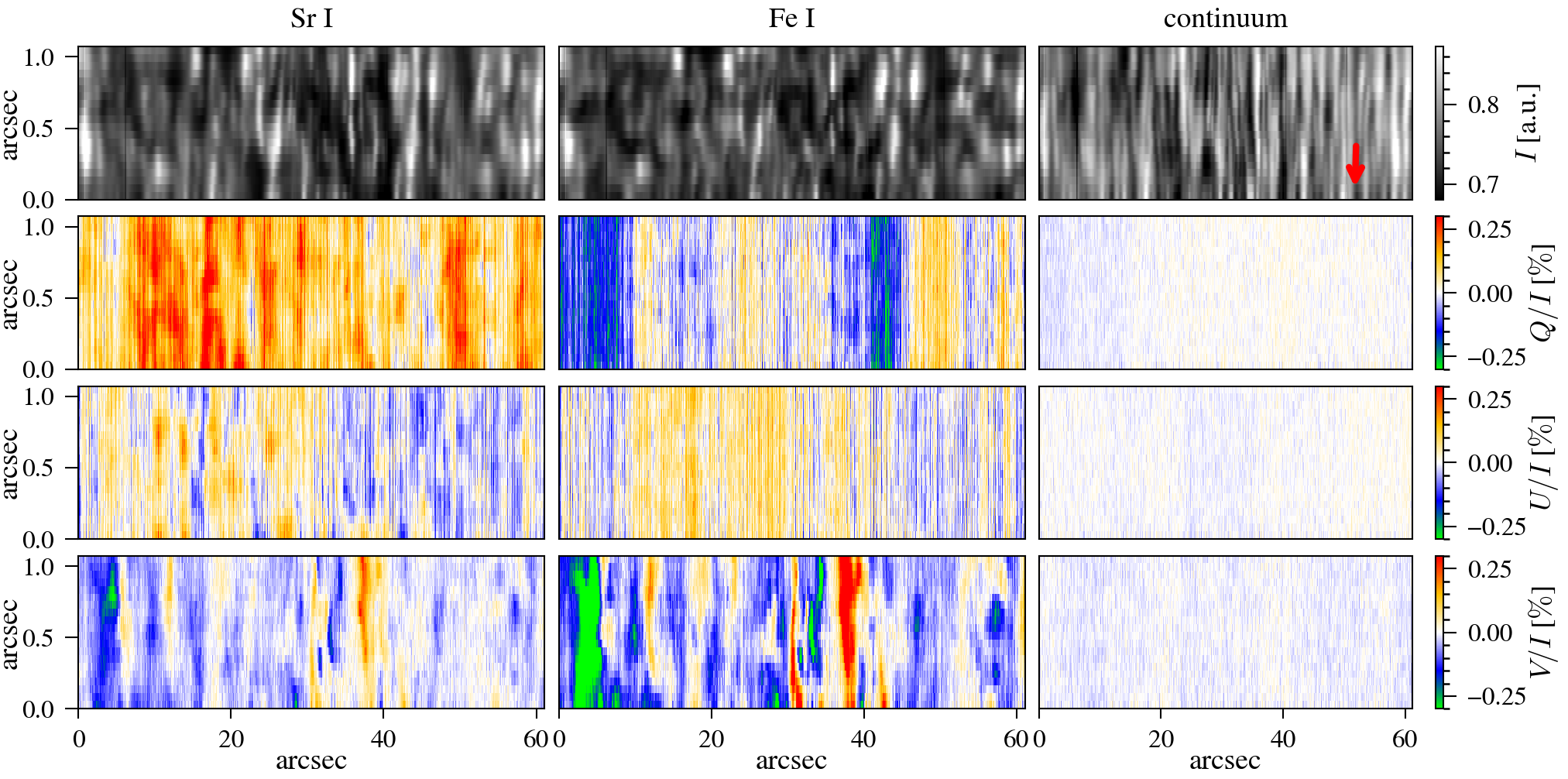}
    \caption{
       Full Stokes maps in Sr~{\sc i}, Fe~{\sc i}, and continuum from ViSP scanning a 61\farcs8$\times$1\farcs1 region at $\mu=0.74$. The data is spatially binned to 0\farcs1$\times$0\farcs1 sampling. Each colorbar applies to all panels left from it, except Stokes $I$, whose range applies only to the continuum. The disk center direction is indicated by the red arrow and the scanning direction is upwards (positive direction of the ordinate). Note that the scale in the scanning direction is much smaller compared to the scale along the slit.
    }
    \label{fig:maps}
\end{figure*}
%
%-----------------------------------------
\subsubsection{Spatial variation of polarization}
\label{sec:spatial_polarization}
%-----------------------------------------
Of the two scans conducted for the position at $\mu=0.74$, we focus our analysis of the spatial variation of the total linear polarization on the one scan with the best image quality.
Figure~\ref{fig:maps} presents the spatially binned Stokes maps of the quiet Sun region scanned by ViSP for Sr~{\sc i}, Fe~{\sc i}, and the continuum, respectively. 
Although the view is projected at $\mu = 0.74$, solar granulation remains clearly visible in the intensity continuum image.
A strong resemblance between Sr~{\sc i} and Fe~{\sc i} is noticeable in the intensity images, indicating that both lines probe similar atmospheric depths. 
The Stokes $V/I$ maps also appear similar in shape.
Differences in the amplitude can be attributed to different Landé factors. 
If the linear polarization signals in Sr~{\sc i} were solely due to the Zeeman effect or systematic errors, we would expect its linear polarization pattern to closely follow that of Fe~{\sc i}. 
However, both the signal amplitudes and structures differ significantly, indicating that the Sr~{\sc i} polarization is not merely a result of Zeeman signals or residual data processing artifacts. 

Remarkable fine structure is visible below the spatial scales of one arcsec, in $Q/I$ as well as in $U/I$. 
There appears to be a slight preference for positive $U/I$ values on the left and negative values on the right side of the slit center. 
This pattern may indicate a residual from an imperfect rotation between $Q/I$ and $U/I$, as discussed in Section~\ref{sec:reduction}. 
This effect is more apparent in the Sr~{\sc i} line, where the scattering polarization tends to produce a predominantly positive $Q/I$ signal. 
An imperfect rotation would therefore leak some of this signal into $U/I$, creating an artificial asymmetry. 
In contrast, Fe~{\sc i} does not exhibit the same positive bias in $Q/I$, making such rotation residuals less pronounced. 
The presence of a similar trend in Fe~{\sc i} $U/I$ nonetheless suggests that an additional physical contribution—such as transverse Zeeman signals superimposed on the scattering polarization—might also be at play.
The spatial average $\langle Q/I\rangle$ is 0.12\% in Sr~{\sc i}, whereas for $\langle U/I\rangle$ it is below the noise level of $\sigma_{{\mathrm Sr},b}$. 
Both spatially averaged linear polarization values in Fe~{\sc i} are below the expected noise level. 
The continuum polarization is free from any small-scale structure (e.g., resembling structures found in the intensity image), indicating that seeing-induced cross-talk is below the noise level.

The observed region is magnetically very quiet, as indicated by the low $V/I$ signal in Fe~{\sc i}. 
We remind the reader that this image is taken in the red wing and not in the core of the line (vertical dashed red lines in Figs.~\ref{fig:intensity_spectra} and \ref{fig:clv_q}).
$V/I$ is sensitive to the LOS magnetic field via the longitudinal Zeeman effect and does not exceed 1.5\% in absolute value anywhere in the spectrum for the entire FoV. 
Since both Fe~{\sc i} spectral lines have a larger effective Landé factor than Sr~{\sc i}\footnote{$LS$ coupling assumed, $\bar{g}$ given by \citet{LandiDeglInnocenti1982}: $\bar{g}_{LS}^{\mathrm{Sr}}=1$, $\bar{g}_{LS}^{\mathrm{Fe,1}}=1.25$ and $\bar{g}_{LS}^{\mathrm{Fe,2}}=1.37$.}, they are more sensitive to magnetic fields.
Thus, the low polarization amplitudes in Fe~{\sc i} provide a robust reference for confirming the absence of significant LOS magnetic fields in this region.

As residual rotation between $Q/I$ and $U/I$ cannot be entirely ruled out (see Section~\ref{sec:reduction}), hereafter we focus our analysis to the total linear polarization \mbox{$P_{\mathrm L}$} in Sr~{\sc i} and Fe~{\sc i}, to estimate which parts of the FoV exhibit more linear polarization than others. 
To explore how $P_{\mathrm L}$ relates to granular structures in the photosphere, we examine its spatial distribution. 
Our scanned FoV is too small to apply standard granule segmentation techniques reliably \citep[e.g.,][]{Liu2021,DiazCastillo2022}. 
Instead, we classify granules and intergranular lanes using a combination of continuum intensity and the LOS velocity $v_{\rm los}$. 
The LOS velocity is calculated from the Doppler induced line shifts in Sr~{\sc i}, where the zero is the spatial mean line core position.
Thereby we ignore Doppler shifts due to the solar rotation.
We note that the inferred LOS velocities (also in the core of the Fe~{\sc i} line) appear lower than values commonly reported for high spatial resolution observations. 
This is expected given the achieved spatial resolution of $\sim0\farcs2$ and minor stray-light contributions.
While these effects reduce the apparent amplitude of granular velocities, the exact velocity values are not critical for the classification used here.
Unlike previous studies \citep[e.g.,][]{Malherbe2007,Zeuner2018,Dhara2019}, which relied solely on continuum intensity, by incorporating Doppler velocity information we thereby account for some projection effects and add more robustness to the classification into granules or intergranules.
This is particularly relevant at our off-center viewing angle.
\begin{figure}[ht]
    \includegraphics[width=0.5\textwidth]{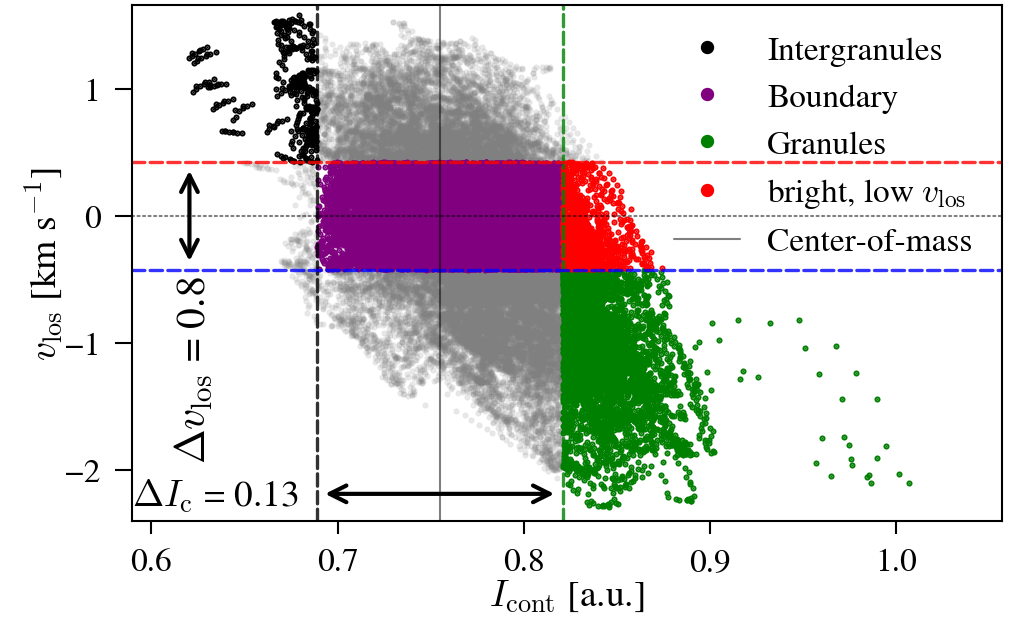}
    \caption{
     Scatter plot of the continuum intensity and LOS velocity estimated from Sr~{\sc i} for $\mu=0.74$. Vertical dashed black and green lines indicate the positions of the intensity thresholds for the intergranules and granules, respectively. Horizontal dashed red and blue lines indicate the LOS velocity thresholds for intergranules and granules, respectively. The choice of the thresholds are given in the main text. 
    }
    \label{fig:scatterplot_ic_vlos}
\end{figure}

\begin{figure}[ht]
    \includegraphics[width=0.5\textwidth]{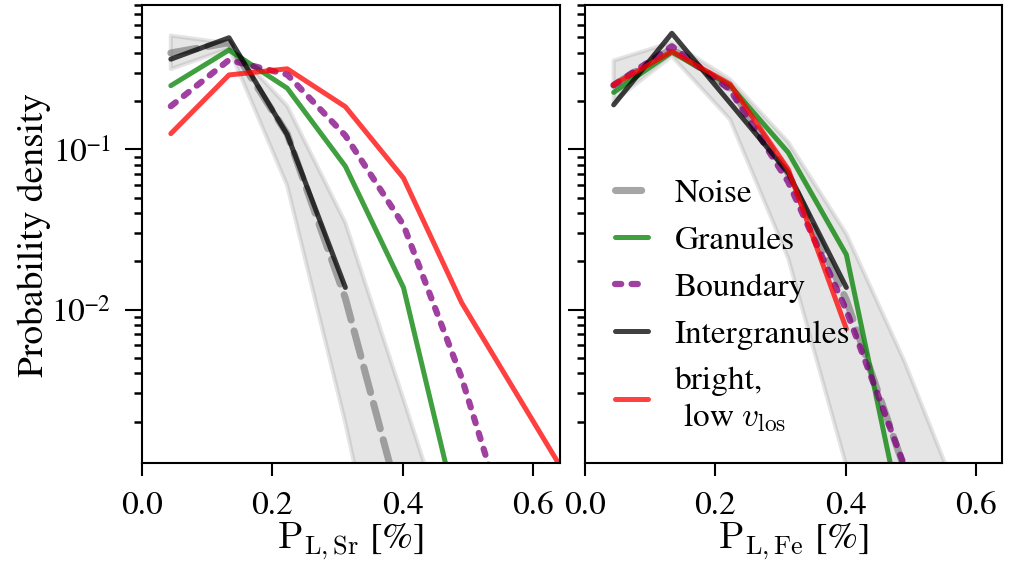}
    \caption{
     Histograms of the total linear polarization of different pixel categories in the Sr~{\sc i} (left) and in the Fe~{\sc i} (right) line cores. The categories are given by thresholds in the continuum intensity and $v_{\rm los}$ (see text for details). The noise distributions (gray area) are the zero-signal reference distributions $P_\mathrm{L, zero}$ (see text for details).
    }
    \label{fig:total_linear_polarization_histogram}
\end{figure}

We emphasize that our classification does not imply that a granular (intergranular) pixel definitely belongs to a granule (intergranule).
Instead, we use the terms granule and intergranule to describe pixels that exhibit characteristic properties: brightness in the continuum intensity and negative LOS velocities for granules, or darkness and positive LOS velocities for intergranules.
Figure~\ref{fig:scatterplot_ic_vlos} shows a scatter plot of continuum intensity versus the LOS velocity for all pixels. 
We apply a threshold-based segmentation using two thresholds for each property; the four thresholds are indicated by dashed lines in the figure. 
The choice of the threshold values will be discussed later.
From this classification, we select three categories of pixels: granular (green), intergranular (black), and a boundary group (purple), which includes pixels that are neither particularly bright/dark nor exhibit significant Doppler
shifts.\footnote{An illustrative 13\arcsec\ subregion from Fig.~\ref{fig:maps} is shown in Appendix~\ref{sec:los_map}.}
Using this classification, we compute normalized histograms (hereafter just histograms) of total linear polarization for each group, as shown in Fig.~\ref{fig:total_linear_polarization_histogram}. 

The central dashed gray line marks the noise calculated with the spatially averaged line depths the zero-signal distribution $P_\mathrm{L, zero}$ explained in the previous section, while the upper and lower limits take into account the range of line depths throughout the FoV. 
This provides an estimate of the minimum and maximum polarization noise levels expected in our data, serving as a reference for interpreting signal significance.
Because binning may distort the noise distribution (see previous section), the histograms are constructed from unbinned data to ensure accurate noise representation.
For Fe~{\sc i}, the total linear polarization in all categories closely follows the estimated noise distribution, indicating no significant polarization signal above noise. 
This agreement confirms that our noise estimation method is consistent with the observed data. 
In particular, it validates the correction factor of two introduced in the previous section, which compensates for the additional broadening of the noise distribution beyond that expected from pure photon statistics.
In contrast, for Sr~{\sc i}, both boundary and granular categories exhibit significant total linear polarization, substantially exceeding the estimated noise, while only the intergranular category shows polarization levels consistent with then noise distribution. 
This reinforces that our estimated noise level is realistic and that the enhanced Sr~{\sc i} signals cannot be attributed to noise misestimation.
We found that the continuum intensity threshold used to define intergranules notably affects the Sr~{\sc i} results. 

We set the lower continuum threshold to include enough intergranular pixels for robust statistics, while avoiding the inclusion of pixels that would shift the polarization probability density of this category toward higher values, thereby keeping it consistent with noise.
Even so, only about 350 pixels meet our intergranular criteria, and therefore, the corresponding histogram (black line) is truncated at the $10^{-2}$ level.
In contrast, the polarization signals in the granule and boundary category are largely insensitive to variations in the upper continuum threshold, suggesting robustness of the total linear polarization above the noise in those categories. 
However, depending on the chosen continuum thresholds, the granule and boundary categories may yield nearly identical histograms.
Sensitivity tests indicate that our findings are weakly dependent on the precise LOS velocity thresholds used.

Interestingly, the highest total linear polarization values are found in pixels that do not fall into any of the three main classifications (intergranule, boundary, or granule). 
Specifically, bright pixels with low LOS velocities—highlighted in red in Fig.~\ref{fig:scatterplot_ic_vlos}—exhibit significantly stronger total linear polarization than either granules or boundary categories (see red line in Fig.~\ref{fig:total_linear_polarization_histogram}). 
This result is largely insensitive to the specific threshold values chosen.
Nonetheless, we selected thresholds that maximize the total linear polarization signal within this category. 
We refer to this category as bright, low $v_{\mathrm {los}}$ pixels. 
Likely, they correspond to the overturning zone in simplified models of granulation.
This aspect will be discussed in Section~\ref{sec:discussion}.
The spatial mean of the total linear polarization in the granular transition is 1.8 times stronger than in the intergranules.
The remaining intermediate classes were also examined; their polarization distributions show no distinctive behavior, lying between those of interganular and bright, low $v_{\mathrm {los}}$ pixels.
%-----------------------------------------
\subsubsection{S/N distribution}
\label{sec:snr}
%-----------------------------------------
%
\begin{table}
    \centering
    \begin{tabular}{r|ccccc}
  &\multicolumn{4}{c}{  S/N=$P_{\mathrm{L, Sr}, b}/\sigma^{P}_{\mathrm{Sr}, b}$}\\
        & $<$ 2  & 2-3  & 3-4  &  $>$ 4\\
        \hline
    $\langle P_{\mathrm {L, Sr}, b }\rangle \ [\%]$ & 0.17 & 0.27 & 0.37 &  0.47 \\
    $\langle P_{\mathrm {L, Fe}, b }\rangle \ [\%]$ & 0.09 & 0.10 & 0.11 &  0.11 \\
        $f_a \ [\%]$ & 40.0 & 47.0 & 12.8 & 0.25  \\
    \end{tabular}
    \caption{Spatially averaged total linear polarization and covered fractional area at different signal-to-noise levels given by Sr~{\sc i}.}
    \label{tab:pl_snr}
\end{table}
\begin{figure}[ht]
    \includegraphics[width=0.5\textwidth]{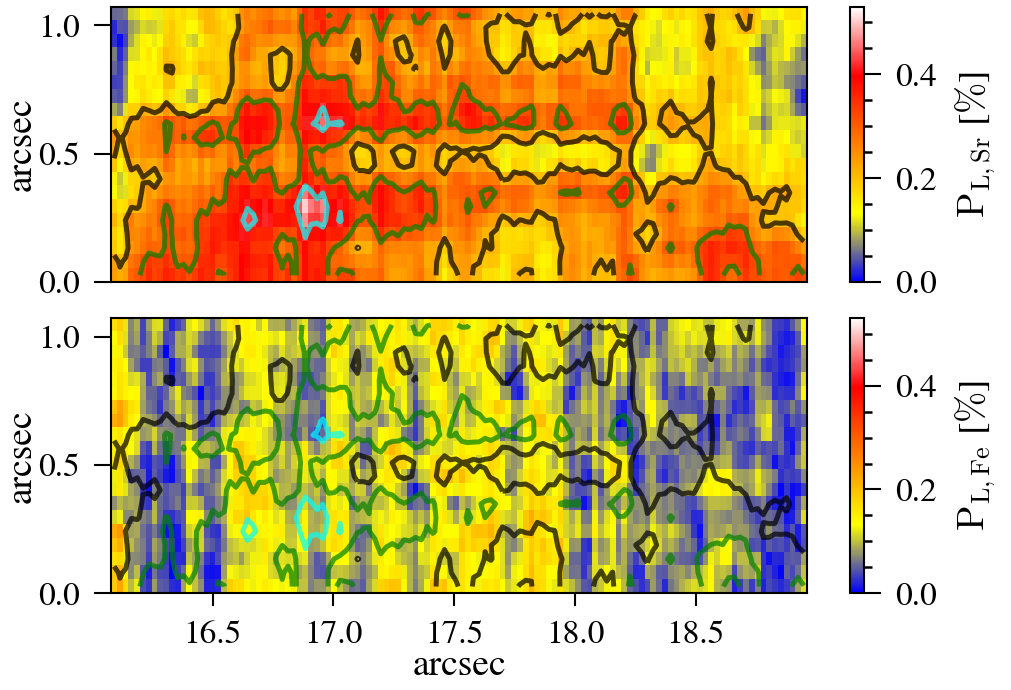}
    \caption{
     Total linear polarization maps for Sr~{\sc i} (upper panel) and Fe~{\sc i} (lower panel) with contours at the S/N=$P_{\mathrm{L, Sr}, b}/\sigma^P_{\mathrm{Sr}, b}$ levels 2, 3, and 4, indicated by the line colors black, green and cyan, respectively, for a detailed region of Fig.~\ref{fig:maps} at $\mu=0.74$.
    }
    \label{fig:total_linear_polarization_snr}
\end{figure}
To assess the spatial scales of total linear polarization at different S/N thresholds, we analyze the total linear polarization in Sr~{\sc i} using S/N=$P_{\mathrm{L, Sr},b}/\sigma^{P}_{{\mathrm Sr}, b}$ levels of 2, 3, and 4. 
A representative subregion of the full map is shown in Fig~\ref{fig:total_linear_polarization_snr}, where contours corresponding to these S/N levels are overlaid on both the Sr~{\sc i} and Fe~{\sc i} total linear polarization maps. 
This zoomed-in view highlights the fine structure of the polarization signals, particularly in Sr~{\sc i}.
We compute the spatial mean of the total linear polarization $\langle P_{\mathrm{L} }\rangle$ between successive S/N levels. 
To ensure the robustness of our results, we also calculate the spatial median, which closely matches the mean -- indicating the distribution is not strongly skewed by outliers. 
The results are summarized in Table~\ref{tab:pl_snr}, which lists the mean total linear polarization amplitudes for both Sr~{\sc i} and Fe~{\sc i}. 

Even at the highest S/N levels, the Fe~{\sc i} structures remain close to the noise level.
This suggests the distinct physical origin of the scattering polarization in Sr~{\sc i}, in contrast to the magnetically-sensitive Zeeman signals that dominate Fe~{\sc i}.
In particular, polarization signals in Sr~{\sc i} cannot be attributed to transverse Zeeman signals (which might be present even in the very quiet Sun), systematic errors, or instrumental cross-talk alone. 

Additionally, we assess the fractional area $f_a$ covered by total linear polarization signals in Sr~{\sc i} at each S/N level, which may serve as a proxy for the spatial filling factor of scattering polarization signals associated with small-scale structures. 
These coverage fractions are listed in Table~\ref{tab:pl_snr}. 
About 40\% of the pixels on the FoV have a S/N of less than 2.
Notably, only 0.25\% of the FoV shows total linear polarization signals with a S/N greater than 4, highlighting the highly localized nature of strongly polarized features.
In terms of spatial extent, these structures are smaller than 0\farcs5 and reach polarization amplitudes of up to 0.42\% (see cyan contours in Fig.~\ref{fig:total_linear_polarization_snr}). 
As expected, the typical size of polarized features decreases with increasing S/N, implying that the strongest signals are associated with the smallest structures.
\citet{DelPinoAleman2018} also found that a reduction of spatial resolution results in lower polarization amplitudes.
These findings emphasize the need for higher S/N across a larger portion of the FoV to enable detailed studies of the spatial distribution of scattering polarization. 
At the same time, preserving high spatial resolution is critical, because the most significant polarization signals are concentrated in relatively small areas.

%%%%%%%%%%%%%%%%%%%%%%%%%%%%%%%%%%%%%%%%%%%%%%%%%%%%%%%%%%%%%%
\section{Discussion and conclusions}
\label{sec:discussion}
%%%%%%%%%%%%%%%%%%%%%%%%%%%%%%%%%%%%%%%%%%%%%%%%%%%%%%%%%%%%%%

Our observations of the Sr~{\sc i} 4607\,\AA\ line for $\mu = 0.74$ show clear spatial structuring in the linear polarization, particularly in $U/I$, with alternating positive and negative values. 
The absence of similar spatial patterns in Fe~{\sc i} suggests that the observed polarization structure in Sr~{\sc i} arises mainly from scattering and possibly the Hanle effect rather than the Zeeman effect or instrumental artifacts.
This spatial structure, consistent with earlier theoretical work \citep{TrujilloBueno2007, DelPinoAleman2018}, results in an average $U/I$ signal close to zero. 
The agreement with these predictions reinforces the role of symmetry-breaking effects in shaping the scattering polarization.

The enhanced spatial resolution of our ViSP observations \citep[down to $\sim$0\farcs2, see][]{Zeuner2025} reveals sub-arcsecond-scale polarization features that remained partially unresolved in earlier studies, such as \citet{Zeuner2020}, where the spatial resolution was approximately twice as coarse. 
These findings provide a valuable observational benchmark for testing radiative transfer calculations in quiet Sun models resulting from three-dimensional magneto-convection simulations.
A key result of this study is the detection of enhanced total linear polarization ($P_{\mathrm{L, Sr}}$) in granular and boundary categories compared to the intergranular category at $\mu = 0.74$. 
In the most extreme intergranular cases (i.e., dark, strongly redshifted, pixels), polarization signals are compatible with the noise level as defined by our criteria. 
This observation supports theoretical predictions \citep{TrujilloBueno2007, DelPinoAleman2018} that the quiet Sun scattering polarization is non-uniform and modulated by the granulation. 
Moreover, this result is compatible with previous spectrograph observations.
These observations \citep{Malherbe2007,Bianda2018,Dhara2019} similarly reported positive correlations between $Q/I$ and continuum intensity, independent of the limb distance with $\mu>0.2$. 
They interpreted this result as enhanced scattering polarization in granules relative to intergranules, likely due to stronger magnetic depolarization in the latter regions via the Hanle effect.
On the other hand, the opposite trend has been found in filtergraph observations at $\mu=0.6$ \citep[][]{Zeuner2018}.
They found an anti-correlation of $Q/I$ with the continuum intensity.
Rigorously comparing our results with such filtergraph observations would require taking into account a sufficient spatial and spectral degradation of the polarized spectra, which are currently limited by the systematic errors. 

For our observation, the suppression of $P_{\mathrm {L, Sr}}$ in the intergranular group may be interpreted in two ways: either intergranules are weakly polarized due to a higher degree of axial symmetry, or the signals are depolarized via the Hanle effect.
The first explanation, that intergranules are more axially symmetric, is not straightforward. 
If the identified structures are indeed intergranular lanes, one would typically expect them to exhibit less axial symmetry than granules. 
This is because plasma conditions are expected to change more abruptly across the intergranule (i.e., toward the neighboring granule) than along it. 
Indeed, at disk center, intergranular lanes often show enhanced scattering polarization precisely due to this increased asymmetry \citep{TrujilloBueno2007,DelPinoAleman2018}. 
Therefore, attributing the lower polarization to a higher degree of axial symmetry would require a more detailed physical justification, which is currently lacking. 
In contrast, the Hanle depolarization hypothesis offers a more consistent explanation within our current understanding, especially given the known sensitivity of Sr~{\sc i} to weak, unresolved magnetic fields.
However, we cannot rule out that both explanations are present simultaneously.
Disentangling these possibilities with a definitive interpretation of which is the dominant physical process which causes this difference would require a careful comparison with simulations including full 3D polarized radiative transfer in a realistic magnetohydrodynamic model, like those of \citet{DelPinoAleman2018}. 
Such simulations can determine whether magnetic depolarization is expected in pixels matching our intergranular definition.
However, such a study is beyond the scope of this work. 

An additional complication for interpretation is that, away from disk center, the classification into granules and intergranules based on the continuum intensity alone can be compromised by projection effects.
Here, to improve the classification, we combined continuum intensity and Doppler velocity criteria. 
This dual-criteria method better accounts for projection effects closer to the limb, although foreshortening still introduces uncertainties, because even granular centers may appear red-shifted due to the projection of vertical flows \citep{Oba2020}, complicating the interpretation.
Ultimately, more reliable classifications will require disk-center observations.

Interestingly, we detect the largest $P_{\mathrm {L, Sr}}$ in bright pixels with weak LOS velocities, likely part of the overturning zone of granules. 
This zone at the edge of granules exhibits, on average, 80\% stronger total linear polarization signals than found in the intergranular category.
We speculate that the observed scattering polarization in the photosphere may be most efficiently generated in overturning zones, where strong horizontal and vertical velocity gradients coexist. 
The symmetry breaking caused by such velocity structures provides a physical mechanism for enhancing the scattering polarization signals. 
The role of horizontal velocity shear in scattering polarization generation has been emphasized by \citet{DelPinoAleman2018}, and the formation of these gradients has been linked to baroclinic vorticity generation in granular boundaries, as discussed by \citet{Nordlund2009}.

Our noise analysis, based on constructing a zero-signal distribution, indicates that less than 1\,\% of the FoV contains $P_{\mathrm{L, Sr}, b}$ signals with S/N > 4 in Sr~{\sc i}, while less than half the pixels have S/N < 2. 
While this is sufficient to reveal fine structure, it also shows that a large portion of the data is still noise-dominated.
The strongest $P_{\mathrm{L, Sr}, b}$ signals are confined to sub-arcsecond structures (<0\farcs5) with amplitudes below 0.5\%. 
These compact features occupy less than 0.2\% of the FoV, underscoring the need for instruments that combine high polarimetric sensitivity with excellent spatial and spectral resolution. 
The quoted fractions depend on our choice to evaluate the polarization at a single wavelength point.
Including multiple wavelength samples across the line profile could increase the effective sensitivity and thereby modify these statistics, potentially increasing the fraction of pixels exceeding a given S/N threshold.
However, at high spatial resolution the detailed spectral shape of the scattering polarization signal is not yet well constrained, as it results from a complex interplay of magnetic fields, velocity gradients, anisotropic radiation, and collisions.
Once physically motivated models and inversion techniques become available, combining information from multiple wavelength samples is expected to yield a significant improvement in effective S/N, as demonstrated by \citet{DiazBaso2025}.

Although doubling the integration time and ViSP’s camera duty cycle could improve the S/N by up to a factor of two, our analysis indicates that the current data may not be purely photon-noise limited. 
Residual instrumental effects, such as internal ghosts, likely contribute significantly and must be addressed in future observations to fully leverage ViSP’s polarimetric sensitivity.
Recent optical improvements and testing with ViSP have shown improved stray light suppression and internal ghosting reduction from the initial commissioning data sets.

While some of the limiting factors that affect the current spatial resolution of ViSP can be mitigated, such as image jitter induced by vibrations due to laboratory instruments, others, like solar dynamics during exposure, pose fundamental limits. 
Nonetheless, the performance of ViSP demonstrates the feasibility of detecting weak scattering polarization signals in the quiet Sun and motivates further improvements. 
In particular, observations at disk center will help resolve ambiguities in the granule-intergranule classification, minimize foreshortening effects, and provide a more direct view of the vertical plasma dynamics. 
The combination of ViSP data with high-cadence context imaging and simulations promises to significantly advance our understanding of the coupling between plasma flows, tangled magnetic fields, and scattered radiation in the lower solar atmosphere.

%%%%%%%%%%%%%%%%%%%%%%%%%%%%%%%%%%%%%%%%%%%%%%%%%%%%%%%%%%%%%%
\begin{acknowledgements}
We express our gratitude to the referee for providing valuable suggestions that have enhanced the clarity of the paper. 
We thank Thomas Schad for helpful discussions.
F.Z. and L.B. acknowledge funding from the Swiss National Science Foundation under grant numbers PZ00P2\_215963 and 200021\_231308, respectively. T.P.A.'s participation in the publication is part of the Project RYC2021-034006-I, funded by MICIN/AEI/10.13039/501100011033, and the European Union “NextGenerationEU”/RTRP.
T.P.A. and J.T.B. acknowledge support from the Agencia Estatal de Investigaci\'on del Ministerio de Ciencia, Innovación y Universidades (MCIU/AEI) under grant
``Polarimetric Inference of Magnetic Fields'' and the European Regional Development Fund (ERDF) with reference PID2022-136563NB-I00/10.13039/501100011033.
E.A.B. acknowledges financial support from the European Research Council (ERC) through the Synergy grant No. 810218 (``The Whole Sun'' ERC-2018-SyG). 
Work by R.C. was supported by the National Center for Atmospheric Research, which is a major facility sponsored by the National Science Foundation (NSF) under Cooperative Agreement No.~1852977.
This research has made use of NASA's Astrophysics Data System Bibliographic Services. 
The research reported herein is based in part on data collected with the Daniel K. Inouye Solar Telescope (DKIST), a facility of the National Solar Observatory (NSO). The NSO is managed by the Association of Universities for Research in Astronomy, Inc., and funded by the National Science Foundation. Any opinions, findings, and conclusions or recommendations expressed in this publication are those of the authors and do not necessarily reflect the views of the National Science Foundation or the Association of Universities for Research in Astronomy, Inc. DKIST is located on land of spiritual and cultural significance to Native Hawaiian people. The use of this important site to further scientific knowledge is done with appreciation and respect. The observational DKIST data used during this research is openly available from the \href{https://dkist.data.nso.edu}{DKIST Data Center Archive} under the proposal identifier pid$\_2\_70$.
\\
\textbf{Facilities}: {DKIST \citep{Rimmele2022}.}
\\
\textbf{Software}: Astropy \citep{astropy}, Matplotlib \citep{matplotlib}, Numpy \citep{numpy}, SciPy \citep{scipy}, SunPy \citep{sunpy}.

\end{acknowledgements}

%%%%%%%%%%%%%%%%%%%%%%%%%%%%%%%%%%%%%%%%%%%%%%%%%%%%%%%%%%%%%%
% References
%%%%%%%%%%%%%%%%%%%%%%%%%%%%%%%%%%%%%%%%%%%%%%%%%%%%%%%%%%%%%%

\bibliographystyle{bibtex/aa}
\bibliography{bibtex/bib}

%%%%%%%%%%%%%%%%%%%%%%%%%%%%%%%%%%%%%%%%%%%%%%%%%%%%%%%%%%%%%%
% Appendix
%%%%%%%%%%%%%%%%%%%%%%%%%%%%%%%%%%%%%%%%%%%%%%%%%%%%%%%%%%%%%%

\begin{appendix}

%-----------------------------------------
\section{VBI context image at $\mu=0.74$}
\label{sec:vbi}
%-----------------------------------------
%
\begin{figure}[ht]
    \includegraphics[width=0.5\textwidth]{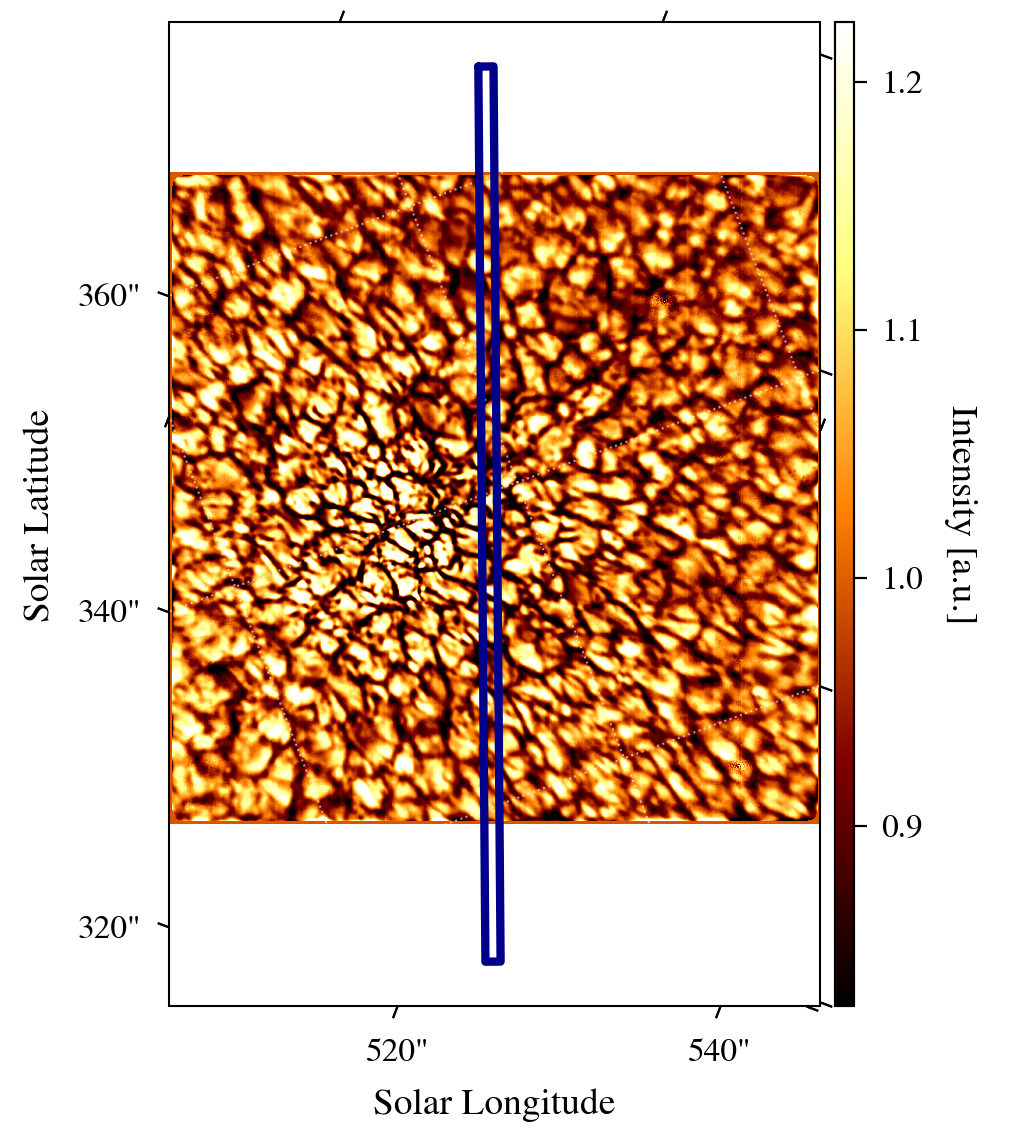}
    \caption{
     Co-temporal VBI context image at 4500 \AA\ corresponding to the data set at $\mu = 0.74$, averaged over all frames belonging to the analyzed scan (10 min). The region scanned by ViSP is indicated in blue. Due to uncertainties in both the VBI coordinate accuracy and the co-alignment between VBI and ViSP, the indicated region is only approximate and intended for illustrative purposes.
    }
    \label{fig:vbi}
\end{figure}

\section{Individual limb scans in Sr~{\sc i}}
\label{sec:appendixA}
\begin{figure*}[ht]
    \includegraphics[width=17cm]{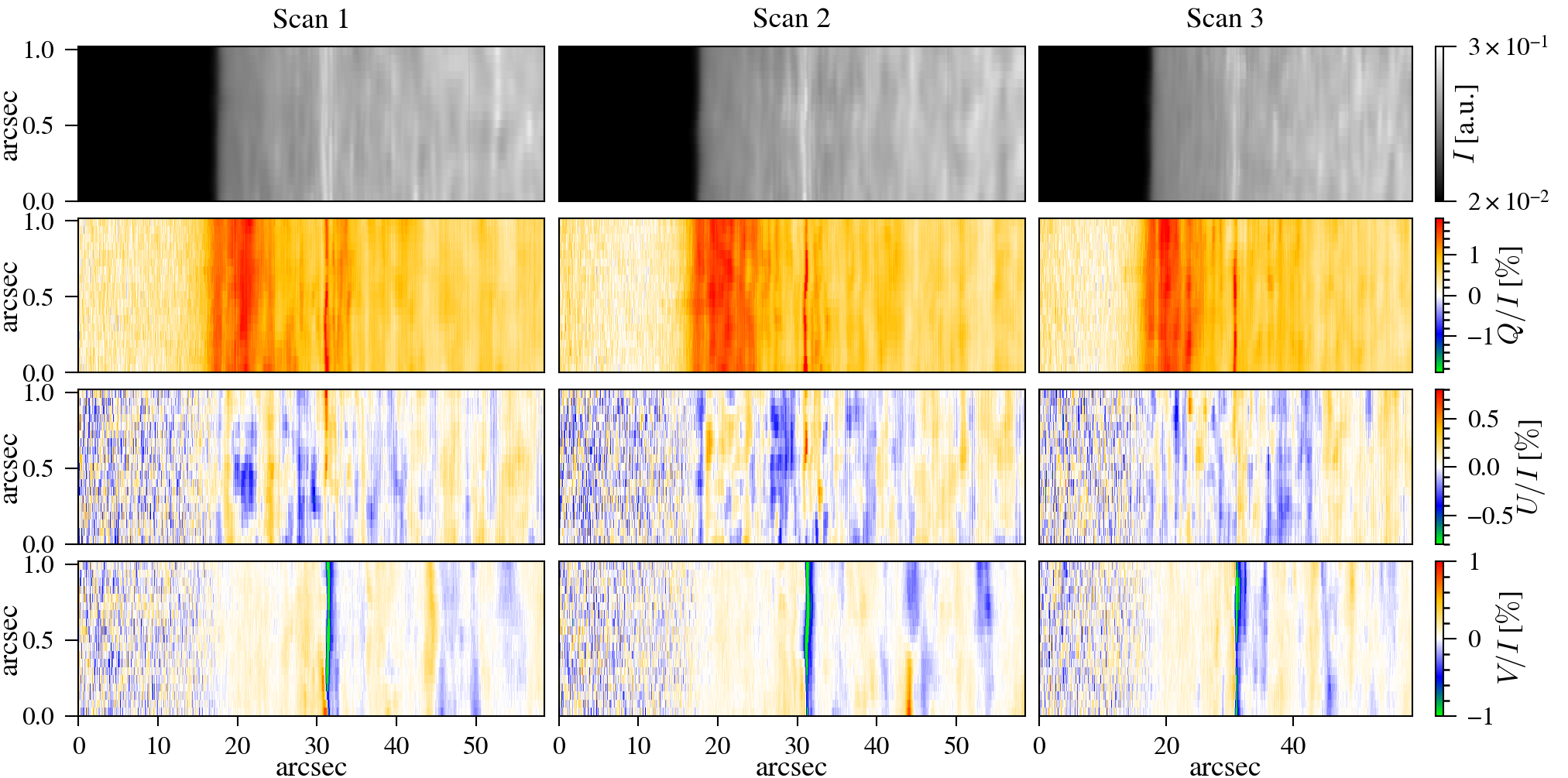}
    \caption{
    Full Stokes maps in Sr~{\sc i} for all three individual scans (from left to right, chronologically) spatially binned to $0{\,}.{\!\!}{\arcsec}1$ $\times$ $0{\,}.{\!\!}{\arcsec}1$ sampling. The intensity and linear polarization is plotted for spectral line core wavelength positions, while the circular polarization is at a small wavelength offset in the red wing. 
     Each Stokes parameter has the same scale shown on the right. 
    }
    \label{fig_app:sr_limb_images}
\end{figure*}

%-----------------------------------------
\section{LoS velocity map at $\mu=0.74$}
\label{sec:los_map}
%-----------------------------------------
%
\begin{figure}[ht]
    \includegraphics[width=0.5\textwidth]{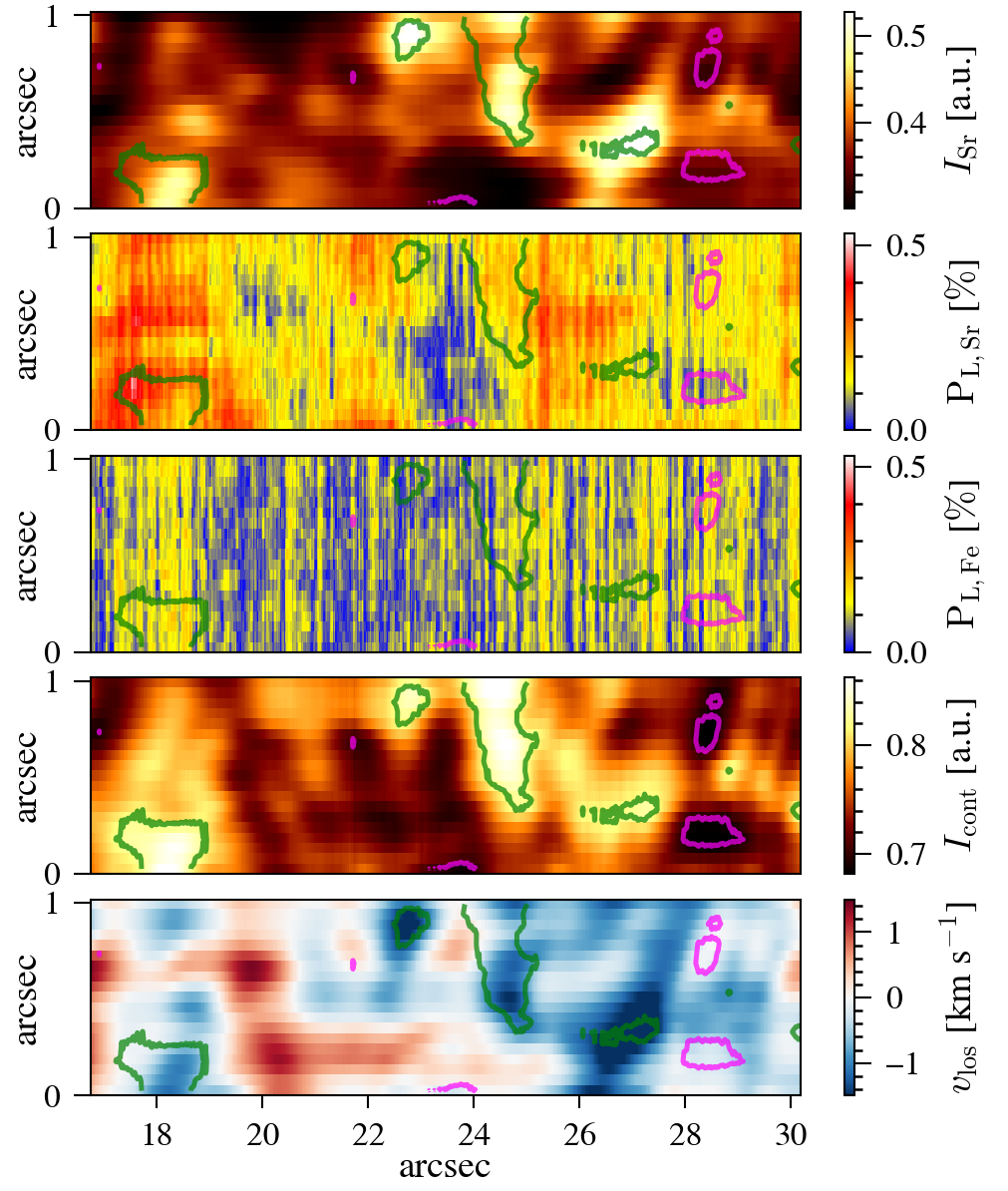}
    \caption{
     Intensity in Sr~{\sc i} (top) and total linear polarization maps (as explained in the main text) in Sr~{\sc i}, in Fe~{\sc i} (second and third panel, respectively), continuum intensity (second last panel) and line-of-sight velocity (bottom panel) at $\mu=0.74$. Note the displayed region is a cut-out from the full scan presented in Fig.~\ref{fig:maps}, covering 13\arcsec. Data is spatially binned to $0{\,}.{\!\!}{\arcsec}1$ $\times$ $0{\,}.{\!\!}{\arcsec}1$ sampling. Contour colors are corresponding to the continuum intensity thresholds given in Fig.~\ref{fig:scatterplot_ic_vlos}, separating dark (magenta lines) and bright (green lines) pixels.
    }
    \label{fig:total_linear_polarization_image}
\end{figure}

\end{appendix}

\end{document}